\documentclass[article]{elsarticle}

\usepackage{lineno,hyperref}
\modulolinenumbers[5]

\usepackage{amsmath,amssymb}
\usepackage{upgreek}
\usepackage{xcolor}
\usepackage{lineno}

\graphicspath{{fig/}}
\usepackage{multirow}

\journal{Nuclear Instruments and Methods in Physics Research A}

\begin{document}

\begin{frontmatter}

\title{High performance picosecond- and micron-level\\ 4D particle tracking with 100\% \mbox{fill-factor} \\Resistive AC-Coupled Silicon Detectors (RSD)}

\author[address1]{M.~Mandurrino\corref{mycorrespondingauthor}}
\cortext[mycorrespondingauthor]{Corresponding author}
\ead{marco.mandurrino@to.infn.it}

\author[address1]{N.~Cartiglia}
\author[address1,address2]{M.~Tornago}
\author[address3]{M.~Ferrero}
\author[address1,address2]{F.~Siviero}
\author[address4,address5]{G.~Paternoster}
\author[address4,address5]{F.~Ficorella}
\author[address4,address5]{M.~Boscardin}
\author[address5,address6]{L.~Pancheri}
\author[address5,address6]{G.-F.~Dalla~Betta}

\address[address1]{INFN, Sezione di Torino, Via P. Giuria, 1 -- 10125 Torino, Italy}
\address[address2]{Universit\'a degli Studi di Torino, Via P. Giuria, 1 -- 10125 Torino, Italy}
\address[address3]{Universit\'a del Piemonte Orientale, Largo Donegani, 2/3 -- 20100 Novara, Italy}
\address[address4]{Fondazione Bruno Kessler, Via Sommarive, 18 -- 38123 Povo (Trento), Italy}
\address[address5]{TIFPA-INFN, Via Sommarive, 14 -- 38123 Povo (Trento), Italy}
\address[address6]{Universit\'a degli Studi di Trento, Via Sommarive, 9 -- 38123 Povo (Trento), Italy}

\begin{abstract}
In this paper we present a complete characterization of the first batch of Resistive AC-Coupled Silicon Detectors, called RSD1, designed at INFN Torino and manufactured by Fondazione Bruno Kessler (FBK) in Trento. With their 100\% \mbox{fill-factor}, RSD represent the new enabling technology for \mbox{high-precision} \mbox{4D-tracking}. Indeed, being based on the \mbox{well-known} charge multiplication mechanism of \mbox{Low-Gain} Avalanche Detectors (LGAD), they benefit from the very good timing performances of such technology together with an unprecedented resolution of the spatial tracking, which allows to reach the \mbox{micron-level} scale in the track reconstruction. This is essentially due to the absence of any segmentation structure between pads (100\% \mbox{fill-factor}) and to other two innovative \mbox{key-features}: the first one is a properly doped \emph{n}$^+$ resistive layer, slowing down the charges just after being multiplied, and the second one is a dielectric layer grown on Silicon, inducing a capacitive coupling on the metal pads deposited on top of the detector. The very good spatial resolution (\mbox{micron-level}) we measured experimentally -- higher than the nominal pad pitch -- comes from the analogical nature of the readout of signals, whose amplitude attenuates from the pad center to its periphery, while the outstanding results in terms of timing (less than 14~ps, even better than standard LGAD) are due to a combination of \mbox{very-fine} pitch, analogical response and charge multiplication.
\end{abstract}


\end{frontmatter}


\section{Introduction}

\begin{figure}[!h]\begin{center}
\includegraphics[width=.7\columnwidth]{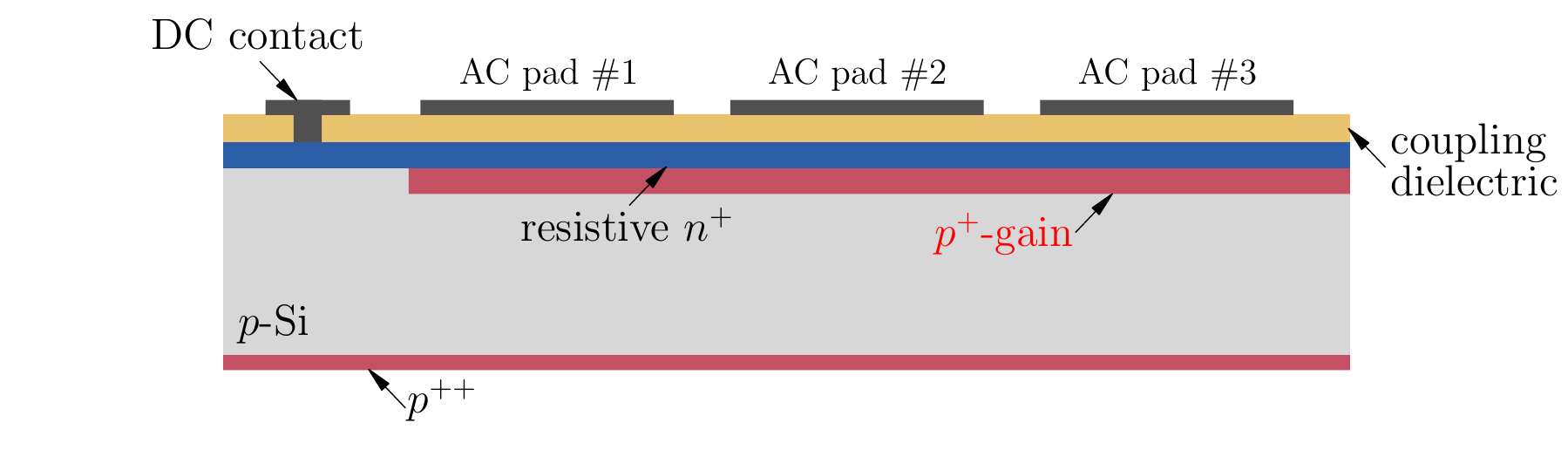}
\caption{Cross-section of an RSD device. In reverse polarization, the secondary charges drift towards their respective electrodes: holes reach the \mbox{\emph{p}$^{++}$-anode} and electrons the \mbox{\emph{n}$^+$-cathode}. Then, the multiplied electrons experience a certain sheet resistance in their discharge path towards the DC contact, which has a characteristic time that allows each metal pad to read a signal. This signal is capacitively induced via the coupling dielectric deposited between Silicon and readout pads.}\label{fig:RSD_cross-section}
\end{center}\end{figure}

Resistive AC-Coupled Silicon Detectors (RSD), an evolution of the \mbox{Low-Gain} Avalanche Detector (LGAD) technology~\cite{2014Pellegrini_NIMA}, are conceived to get rid of all the isolation implants used in traditional segmented trackers to prevent early breakdown at borders and \mbox{short-circuit} phenomena. This means that RSD are no more affected by the unwanted gain loss and, thus, efficiency deterioration in the time tagging of particles coming from the presence of those isolation structures. Moreover, thanks to the use of a continuous multiplication layer, RSD are characterized by a 100\% \mbox{fill-factor} (the ratio between the active area and the total sensor area), an essential ingredient for the \mbox{high-performance} \mbox{4D-tracking}. In this scheme, the space/time reconstruction is obtained thanks to: (\emph{i}) a resistive \mbox{\emph{n}$^+$-type} implant, slowing down the multiplied charges before their discharge -- for a characteristic time which allows the readout and, at the same time, minimizing pile-up effects -- and (\emph{ii}) a dielectric layer responsible for the capacitive induction of charge on the metal pads deposited on top of the sensor (see Figure~\ref{fig:RSD_cross-section}).

The signal generated in an RSD is bipolar, since all the charge induced on the pad has to be restored to zero after the discharge process. Moreover, not only one pad responds to the capacitive induction but a certain cluster of pads, whose dimension depends on the characteristic RC of the detector, is involved in the signal formation. In order to properly design our RSD, all the physical and technological parameters impacting on its working principle have been carefully considered. Among them, we find the optimal dose of \emph{n}$^+$, directly determining the discharge time, the dielectric composition and thickness, which affects both the coupling and the interpad capacitance and influencing the charge sharing and signal amplitude, and finally the gain dose, driving the multiplication factor and, thus, the signal slew rate.

It is straightforward that, having such working scheme and due to the charge sharing mechanism, the RSD readout logic is required to be completely analogical. Anyway, this is not a bottleneck but a true advantage, since it allows to have more information for reconstructing each particle track than the single channel as it occurs in traditional Silicon detectors. In this sense, RSD represent an innovative technology also under the data analysis standpoint.

\section{The RSD1 production}

\begin{figure}[!h]\begin{center}
\includegraphics[width=.83\columnwidth]{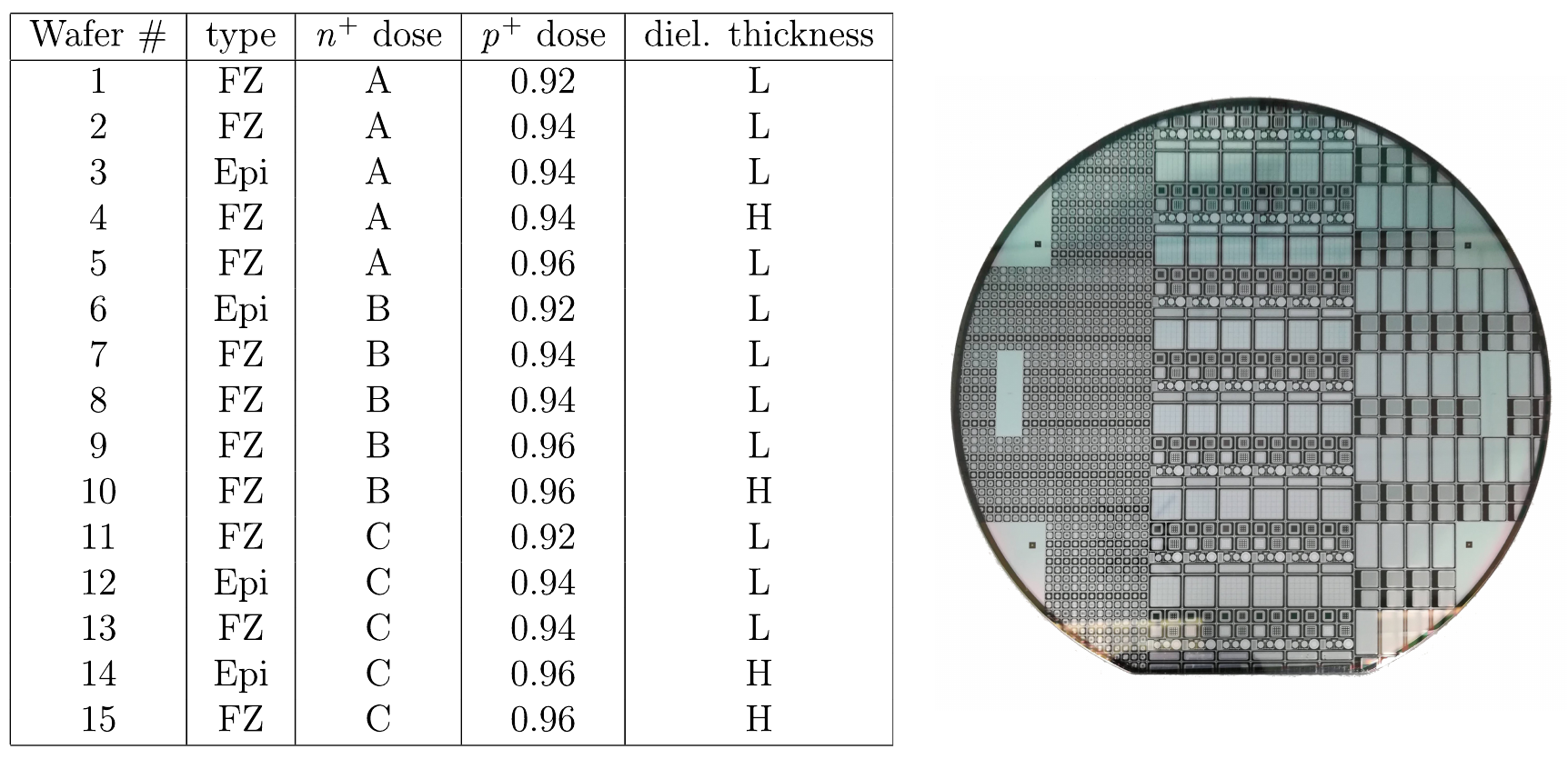}
\caption{Summary of process split parameters in RSD1 batch (left) and final layout of the same production run at FBK (right).}\label{fig:RSD1_batch}
\end{center}\end{figure}

After an intense simulation campaign and an accurate design process, both well described in~\cite{2020Mandurrino_NIMA}, we produced the set of lithographic masks needed for the first batch of detectors RSD1 at FBK. The run is fabricated through the \mbox{step-and-repeat} (stepper) technology on Epitaxial (Epi) and Float Zone (FZ) 6$^{\prime\prime}$ wafers, both with a \mbox{50-$\mu$m-thick} active region. It includes 15 wafers (see the parameters table in Figure~\ref{fig:RSD1_batch}) processed by using different splits of \mbox{$n^+$-resistive} dose (increasing letters means increasing implantation dose) as well as the \mbox{$p^+$-gain} dose (Boron) and the dielectric thickness (where H means high and L stands for low). Notice that, being based on FBK proprietary technologies, the manufacturer is not able to disclose some details about the RSD process.

To test the AC properties of our RSD with respect to the geometrical configuration of pads, different RSD matrices have been designed. As it is shown in the right picture of Figure~\ref{fig:RSD1_batch}, each wafer is divided into three sectors: starting from the leftmost, we have a first sector including small square matrices with different families of pitch from 50 to 300~$\mu$m and different pad size for each of them. In the second sector other square matrices with pitch 500~$\mu$m and 1.3~mm, plus a strip sensor, have been designed. Finally, the last sector includes a 64$\times$64 pixel matrix with \mbox{50~$\mu$m-pitch} designed for CHIPX65, the Torino version of the RD53A ROC~\cite{RD53A_chip}, and other strip modules expressly conceived for the MoVeIT project that aims at characterizing therapeutical particle beams.

When the RSD1 production was released in June 2019, we proceeded to test the run quality as well as all the static characteristics of sensors. To this aim, intensive $I(V)$ and $C(V)$ measurements have been performed on a large subset of structures coming from all the 15 wafers. If the \mbox{current-voltage} curve constitutes an indication of leakage phenomena and gives useful information about the diode breakdown, measuring the capacitance as a function of bias reveals precious details about the multiplication implant, such its doping concentration profile and dose.

\begin{figure}[!h]\begin{center}
\includegraphics[width=\columnwidth]{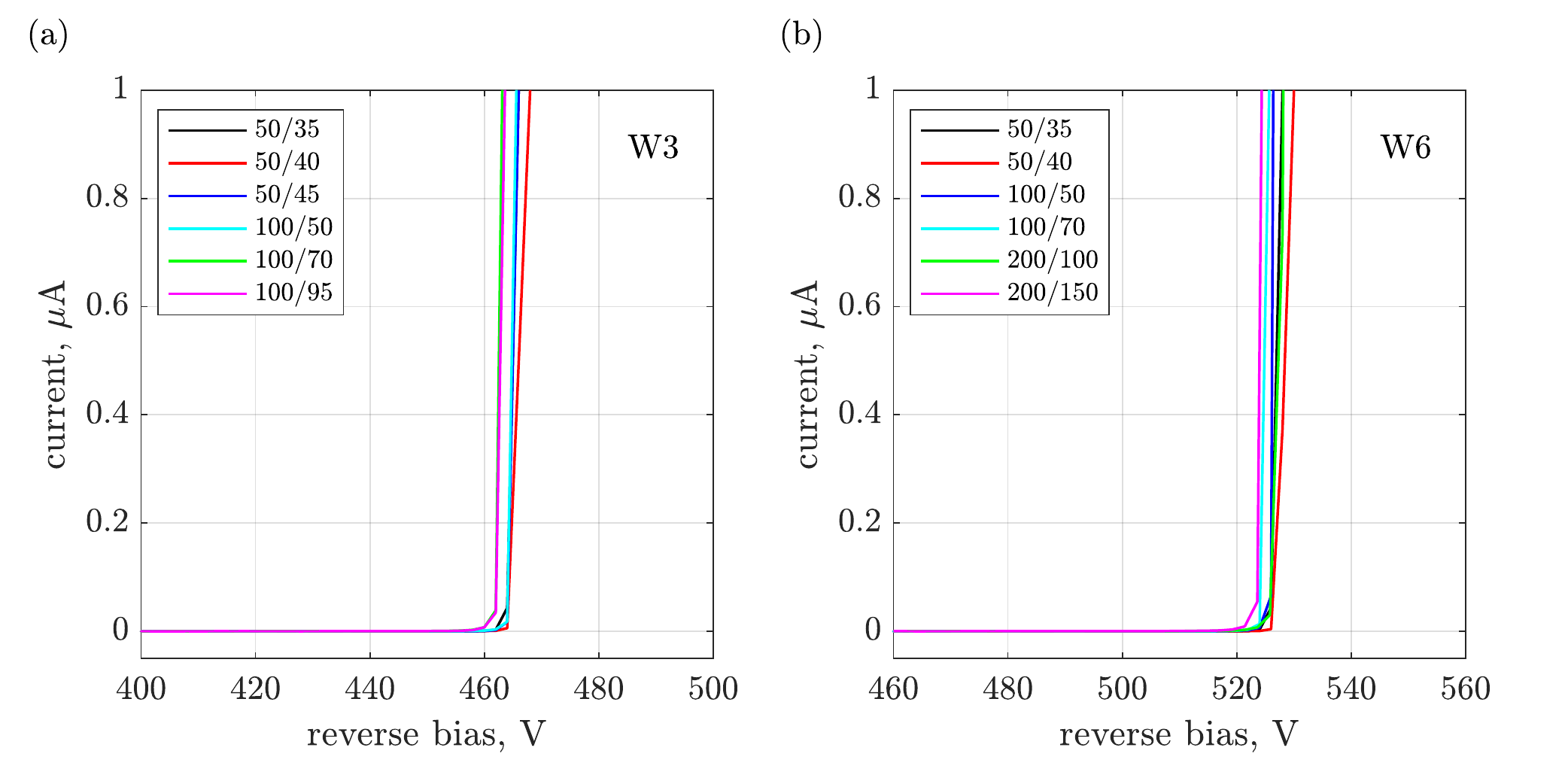}
\caption{Current-voltage characteristics measured in reversely polarized RSD1 with different geometrical configurations of pad \mbox{pitch/side} that come from different positions on W3 (a) and W6 (b) showing a very narrow window of breakdown voltages $V_\text{bd}$, which denotes a very good uniformity of the gain implantation within each wafer.}\label{fig:IV}
\end{center}\end{figure}

Figure~\ref{fig:IV} reports some $I(V)$ characteristics measured for several RSD1 with different pad \mbox{pitch/side} and belonging to wafers 3 and 6. The tested structures come from different coordinates of both wafers and the great homogeneity of the breakdown voltage $V_\text{bd}$ (spread of few Volts) is a demonstration of (\emph{i}) the great uniformity of the multiplication layer implantation within each wafer and (\emph{ii}) the independence of the $V_\text{bd}$ value from the geometry or, in other words, from the depleted volume of the sensor, which changes in the considered structures according to the pitch size.

\begin{figure}[!h]\begin{center}
\includegraphics[width=\columnwidth]{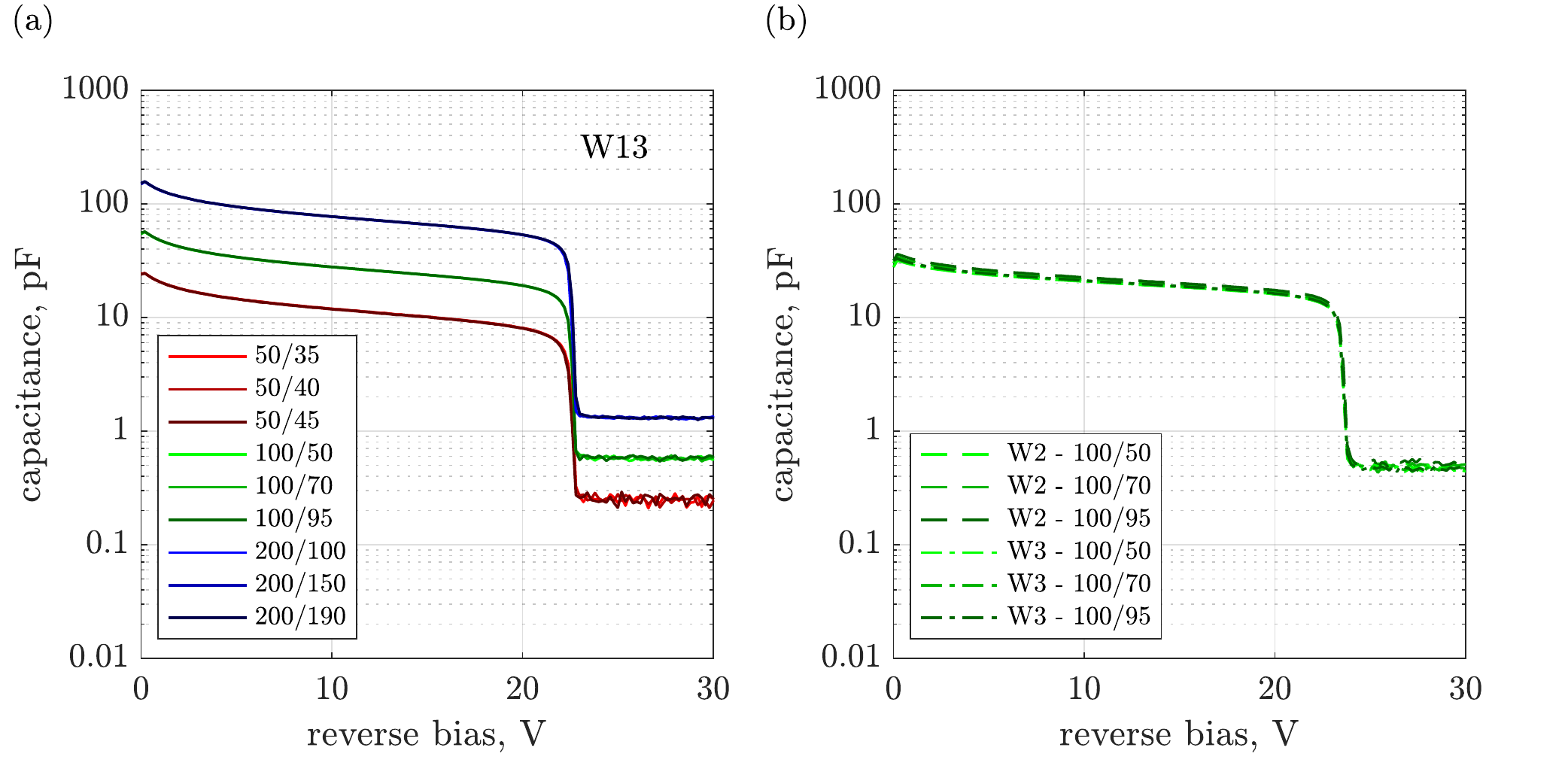}
\caption{Capacitance-voltage characteristics measured in reversely polarized RSD1 with different geometrical configurations of pad \mbox{pitch/side} coming from different positions on W13 (a) or on W2 and W3 (b).}\label{fig:CV1}
\end{center}\end{figure}

Concerning the $C(V)$, we measured the detector capacitance, to test the most important diode implants, as well as both the AC pad and the \mbox{AC-AC} interpad capacitances, in order to have a feedback on the RC characteristics of our RSD1 devices. As an example, Figure~\ref{fig:CV1}(a) shows standard $C(V)$ curves comparing detectors from the same wafer but belonging to three different families of active volumes. The plot demonstrates that the overall $C(V)$ trend and the \mbox{full-depletion} capacitance depend on the pitch (active volume) and that detectors having the same volume have the same $C(V)$, further indicating multiplication implant uniformity in the RSD1 production. The same indications of uniformity, not only within each wafer but also among wafers, comes from Figure~\ref{fig:CV1}(b), which represents several overlapping $C(V)$ characteristics measured in RSD1 with \mbox{100~$\mu$m-pitch} and different pad size taken from different positions of homologous wafers (i.e., having the same gain implant).

\begin{figure}[!h]\begin{center}
\includegraphics[width=\columnwidth]{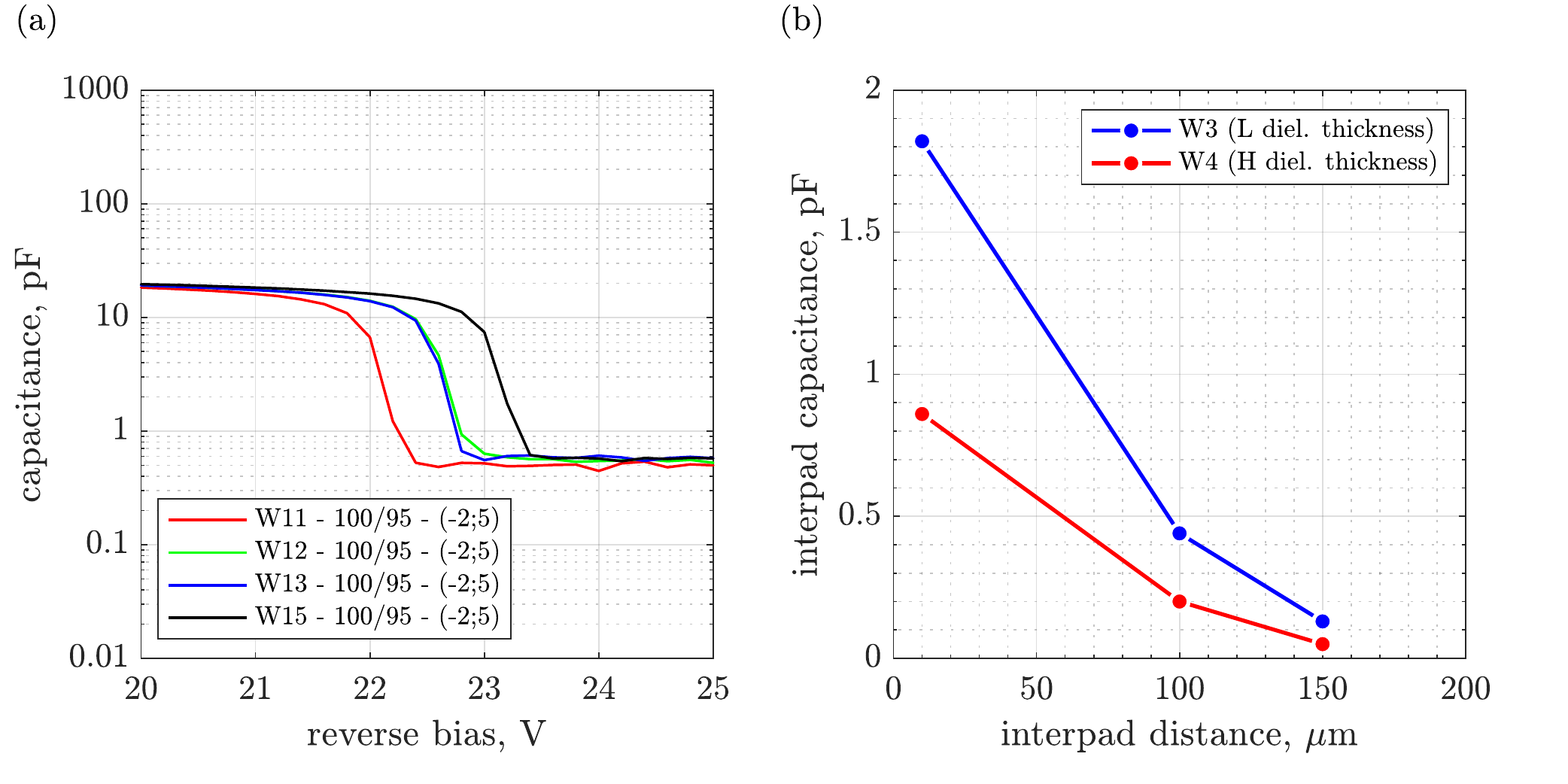}
\caption{Capacitance-voltage characteristics measured in reversely polarized RSD1 with the same pad geometry and coming from the same position -- specified in the legend by \mbox{(-2;5)} -- on different wafers (a) and interpad capacitance as a function of the distance measured in \mbox{300~$\mu$m-pitch} RSD1 devices coming from W3 and W4 (b), that are differing in the dielectric thickness.}\label{fig:CV2}
\end{center}\end{figure}

We also tested the capacitance in devices with the same configuration (pad pitch/side) taken from the same coordinate of wafers W11, W12, W13 and W15 (see panel (a) of Figure~\ref{fig:CV2}) and the results show that both the gain- and the \mbox{full-depletion} voltage properly scale with the $p^+$ dose. Indeed, W12 and W13, characterized by the same gain implant, produce as expected the same $C(V)$, that is different from that of W11 (lower dose) and of W15 (higher dose).

Another important characterization concerns the AC pad and the \mbox{AC-AC} interpad capacitance. The first quantity has been measured by performing a usual $C(V)$ with the $n^+$ contact grounded and the two contacts of our LRC probe connected to the pad (low level) and to the back side (high level) of the device. Then, the AC pad capacitance has been assumed to be the value of C in \mbox{full-depletion} conditions. Measurements are in agreement with theoretical predictions and, in the case of 300/150, 300/200 and 300/290 RSD1 structures are, respectively, 3.40~pF, 5.75~pF and 11.64~pF for W4 and 4.75~pF, 7.90~pF and 14.25~pF for W6. This trend is due to the scaling of the AC pad capacitance with (\emph{i}) the pad area and (\emph{ii}) the dielectric thickness (see W4 and W6 in the table of splits).

Regarding the \mbox{AC-AC} interpad capacitance, we report in Figure~\ref{fig:CV2}(b) the trend as a function of the distance between adjacent pads in \mbox{300~$\mu$m-pitch} RSD1 structures from W3 and W4. Here each data point has been obtained as the full-depletion capacitance of a $C(V)$ curve performed with the probe contacts on the two involved pads, the reverse bias applied on the back side and, again, with the DC contact (resistive implant) grounded. As expected from theoretical predictions, the capacitance scales with the distance. Moreover, it is also inversely proportional to the dielectric thickness, since W3 shows higher values than W4.

\section{Signal characterization}

\begin{figure}[!t]\begin{center}
\includegraphics[width=\columnwidth]{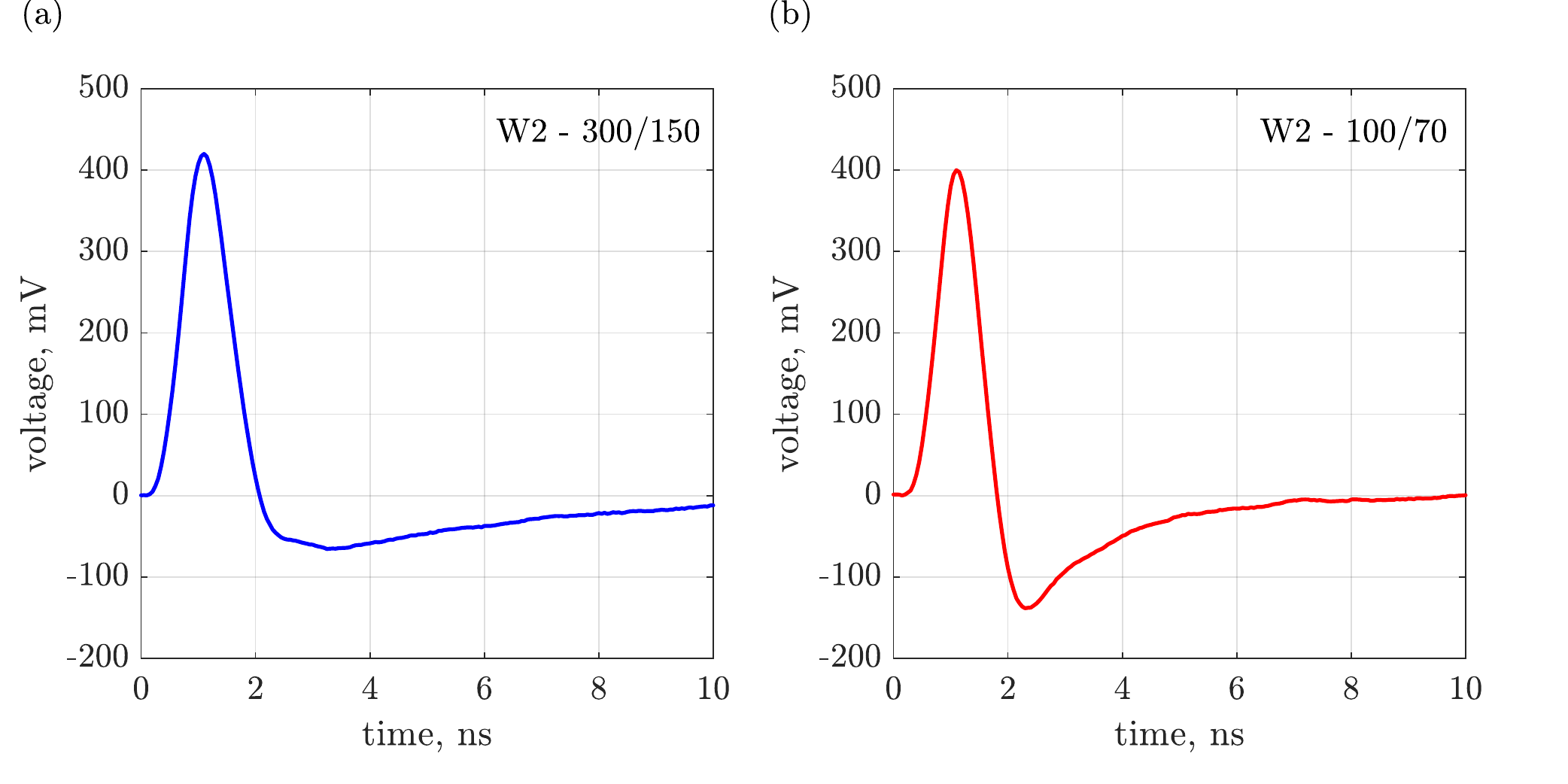}
\caption{Signals (average of several events) acquired with the TCT setup in a 300/150 (a) and 100/70~(b) RSD1 structure from W2.}\label{fig:signals}
\end{center}\end{figure}

Once the RSD1 batch has been characterized by the electrical standpoint, we proceeded to test the signal response of a large set of detectors, exploring all the phase space offered by the split of technological and physical parameters we implemented in this production. The first basic target was to stimulate the production of secondary charges and readout an AC signal from pads. To this aim we used the IR laser (\mbox{$\lambda =$~1060~nm}) of the Transient Current Technique (TCT)~\cite{1993Eremin_NIMA} setup in Torino. Shining the device from the top, the laser penetrates into the Silicon lattice producing secondary charges via ionization. This instantaneously induces a charge on the nearby pads thanks to the \mbox{AC-coupling} and, in the same time, the secondary electrons and holes drift towards their respective electrodes (electrons reach the \mbox{$n^+$-cathode} and holes the \mbox{$p$-anode}) due to the reverse bias. After the multiplication mechanism, the drift of charge carriers drives the shape of the first inductive part of the signal, as stated by the \mbox{Shockley-Ramo} theorem~\cite{1938Shockley_JAP,1939Ramo_IRE}. Then, the electrons populating the resistive cathode start to flow along their discharge path towards the DC contact. In this phase, having no physical collection of electrons, the AC pads affected by the discharge process experience an undershoot in the signal, whose amplitude and duration is determined by the RC characteristics of the RSD, and that has the effect of restoring to zero the overall integrated charge. Some examples of such bipolar nature of the resistive \mbox{AC-coupled} signals can be found in Figure~\ref{fig:signals}, where we reported the pad response in two RSD1 structures from W2 acquired with the TCT: in panel (a) there is a 300/150 detector while in panel (b) a 100/70 structure  is shown.

\begin{figure}[!t]\begin{center}
\includegraphics[width=\columnwidth]{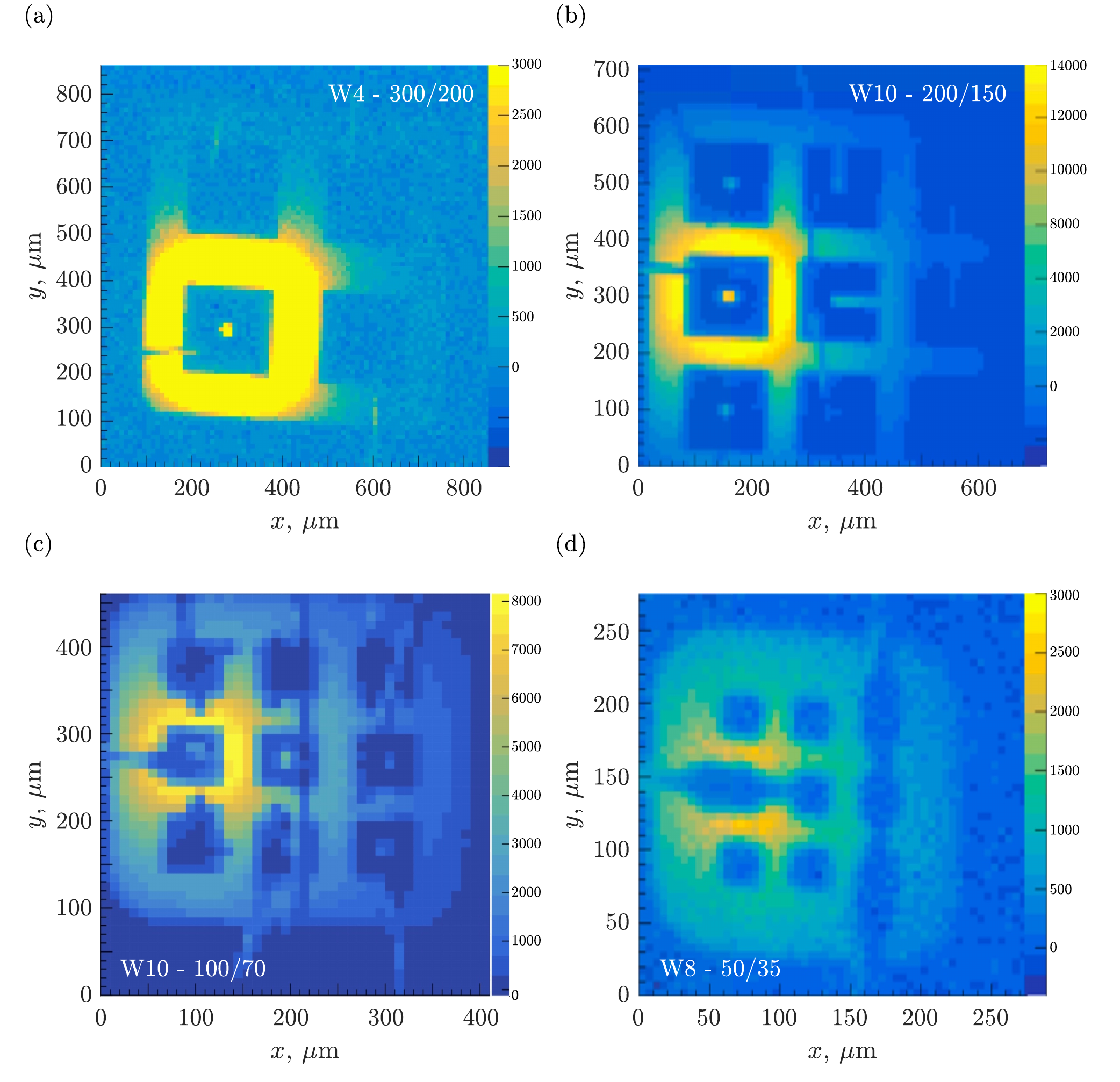}
\caption{2D scans of induced charge in several RSD1 square matrices obtained by integrating the first positive lobe of the AC signals: \mbox{2$\times$2} pad 300/200 from W4 (a); \mbox{3$\times$3} pad 200/150 from W10 (b); \mbox{3$\times$3} pad 100/70 from W10 (c); \mbox{3$\times$3} pad 50/35 from W8 (d).}\label{fig:2Dmaps}
\end{center}\end{figure}

The most relevant aspect coming from the observation of the reported waveforms is that different RC produce different signal shape: indeed, while the sheet resistance and the dielectric thickness are the same in both detectors, the RSD1 shown in panel (a), where the first lobe dominates, has a higher coupling capacitance than the detector in panel (b), where the undershoot is rather prominent, because the pad has a larger area.

Performing a laser scan of all the detector surface rather than shining into a single shot, as in the previous case, we can obtain more information about the overall properties of our RSD1 devices. In particular, it is possible to make a 2D reconstruction of their response if we adequately integrate the first lobe of the signal and assign the resulting charge to its corresponding coordinate, which is given by the TCT setup during the scan. We acquired such data for several detectors of the RSD1 batch and reported some of them in Figure~\ref{fig:2Dmaps} as 2D intensity maps of charge (see Ref.~\cite{2019Mandurrino_EDL} for further details).

We already demonstrated that the RSD1 production represents a first evidence of \mbox{very-fine} pitch detectors with 100\% \mbox{fill-factor}~\cite{2019Mandurrino_EDL}. Here we just want to limit our observations to some important key aspects in view of the \mbox{4D-tracking}. As one may see in the maps, each AC pad is surrounded by an active area for the charge induction. In general, it can be noticed that the charge is not constant around the pad but follows a certain distribution, decreasing radially from the center to the periphery of such area.

\begin{figure}[!h]\begin{center}
\includegraphics[width=\columnwidth]{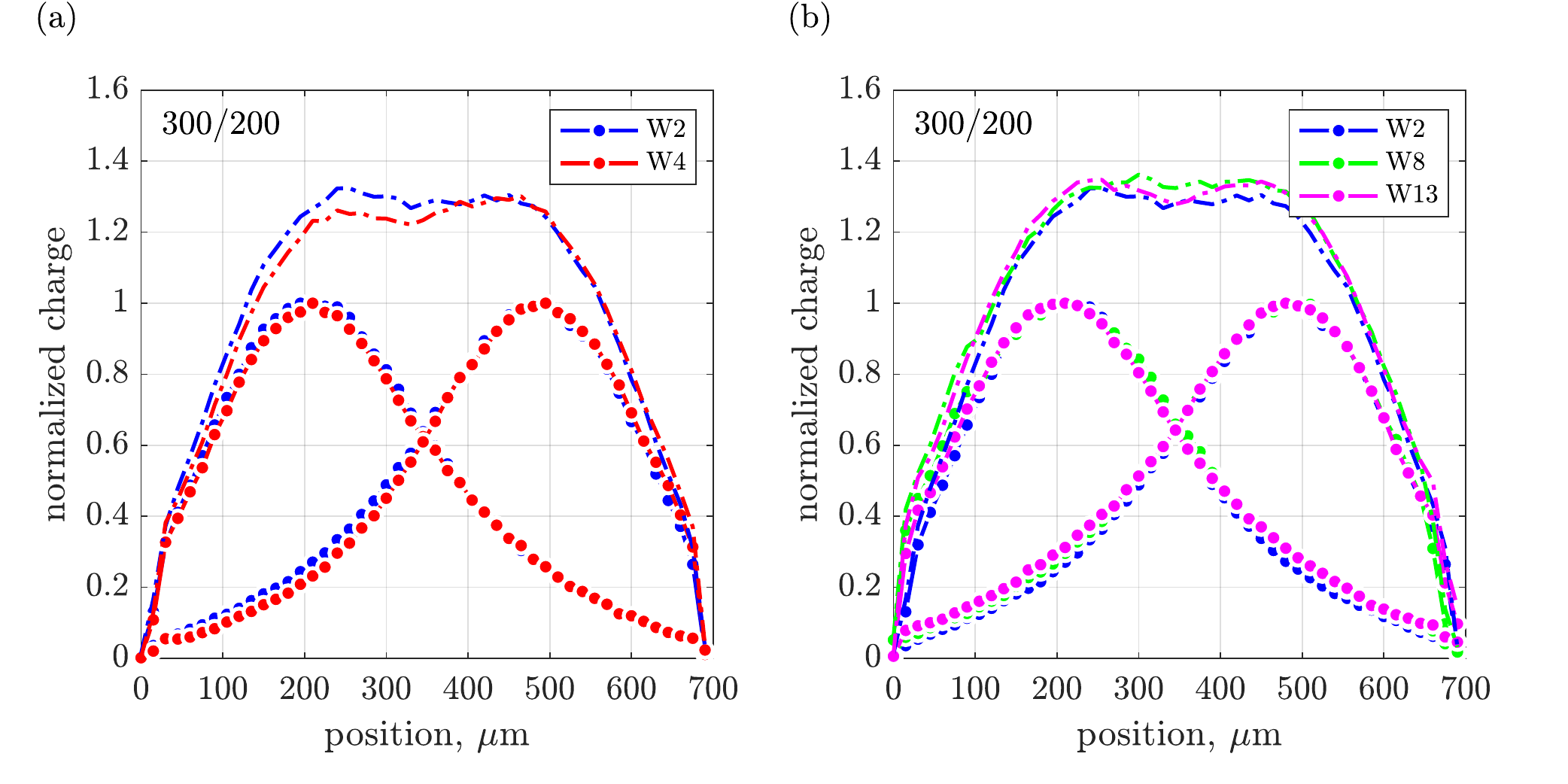}
\caption{Charge projection along a straight line close to two pads in 2$\times$2 300/200 RSD1 structures taken from several wafers differing in the dielectric thickness (a) and $n^+$~dose (b). The \mbox{dot-dashed} curves represent the sum of charges coming from the two pads.}\label{fig:ch_proj}
\end{center}\end{figure}

This can be observed even better by representing the amount of induced charge as a function of the position along a straight line passing close to two neighboring pads, as shown in Figure~\ref{fig:ch_proj}. The charge projections reported there for the same RSD1 structure but coming from the same coordinate of different wafers demonstrate, first, that the 100\% \mbox{fill-factor} requirement is fulfilled, since the sum of charges readout by the two pads (\mbox{dot-dashed} curves) uniformly covers the interpad area in all the wafers. Secondly, the left- and \mbox{right-pad} profiles crosses at \mbox{$\sim$64\%} of the total charge, whatever the dielectric thickness or the $n^+$ dose (see, respectively, panel (a) and (b) of Figure~\ref{fig:ch_proj}). Moreover, we also noticed that this number is sensitive, when choosing a given pitch, to the detector segmentation and that it scales up to \mbox{$\sim$70\%} with the decreasing interpad distance. This is probably due to the \mbox{AC-AC} capacitance, that affects the induction of charge by introducing a further coupling term acting between neighboring pads. As a result, RSD1 samples with a pad pitch/side ratio close to 1 show a sum of charges even more uniform than what we have shown in Figure~\ref{fig:ch_proj}.

\begin{figure}[!h]\begin{center}
\includegraphics[width=\columnwidth]{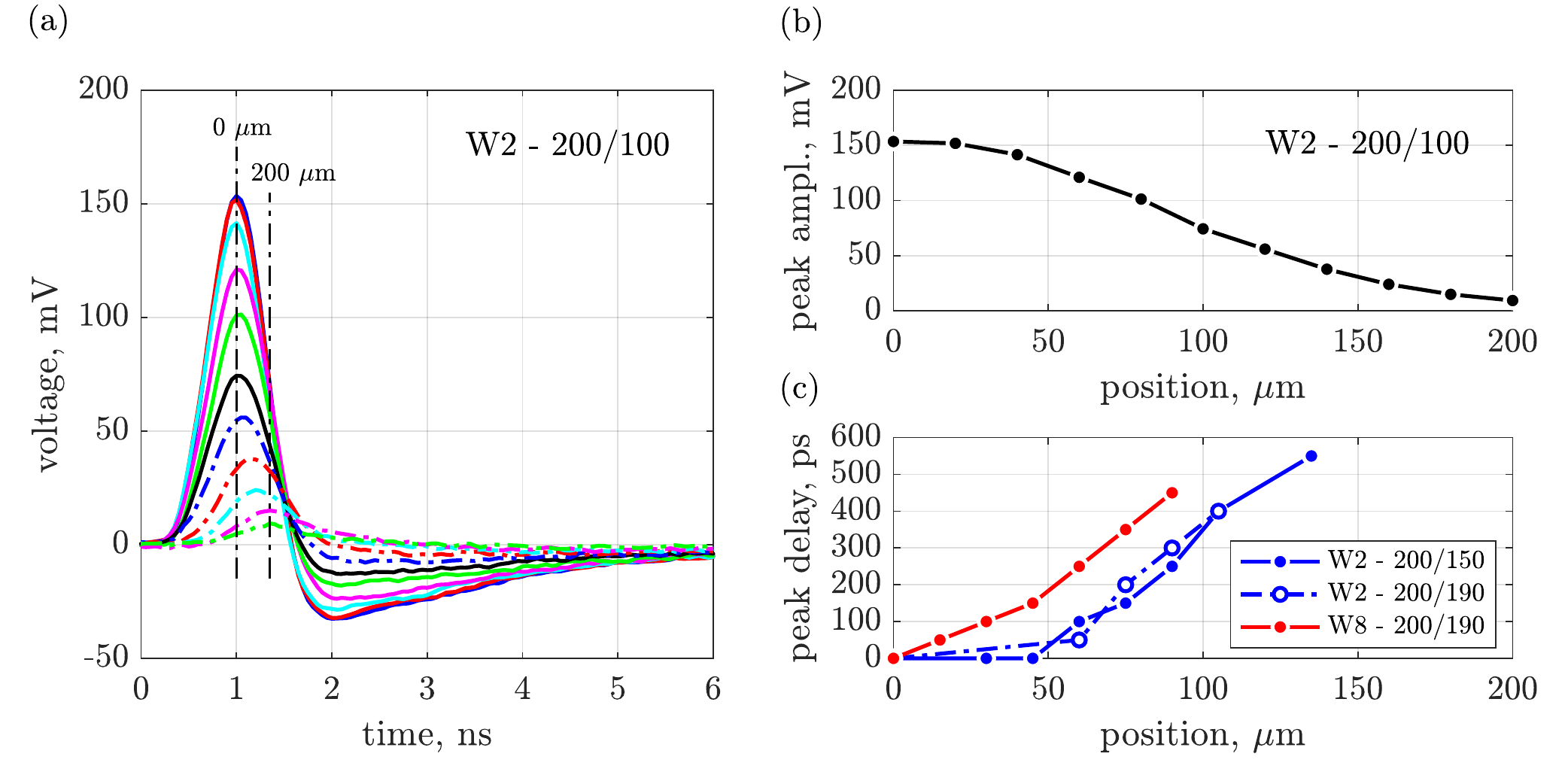}
\caption{TCT measurements of some 3$\times$3 pad \mbox{200~$\mu$m-pitch} RSD1 devices taken during a complete linear scan of the pulsed IR laser between the centers of two adjacent pads, including: (a) a set of 11 waveforms (averages of several hundreds acquisitions) with 20~$\mu$m steps each; (b) the trend of the first lobe peak amplitude as a function of the distance from the pad center for the same structure; (c) delay of the first lobe peak as a function of the distance measured in three different detectors.}\label{fig:signal_delay}
\end{center}\end{figure}

The position where charges are created or, even better, the distance from it to the pad center not only determines the number of induced charges. We observed that also the signal shape significantly changes. In order to deeply investigate this aspect, we predisposed the presence of slits or holes of \mbox{no-metal} in some pads, allowing to make a laser scan between them.

Panels (a) and (b) of Figure~\ref{fig:signal_delay} show, respectively, the waveforms and the maximum signal amplitude as a function of the distance in a \mbox{200~$\mu$m-pitch} RSD1 device during a scan with the TCT laser between the centers of two adjacent pads performed through the non metallized windows described above. As a first observation, one may see that the signal attenuates while the laser spot moves away from the pad center. An effect involving the first lobe as well as the subsequent discharge. This phenomenon justifies the trend of induced charges we observed in the 2D scans and in the projections. Moreover, the second most important aspect is what we condensed in panel (c), i.e. that also the peak time changes as a function of the position. Here we reported three cases, coming from a \mbox{200/100} RSD1 belonging to W2 and a \mbox{200/190} to W2 and W8, from which it results that the peak shift, or delay, with distance holds whatever the geometry or the wafer we choose.

\section{RSD and 4D-tracking}

In this section our tests on the detectors response, and in particular the signal properties just reported, will become of primary importance in view of using RSD as a \mbox{very-high} performance detector for 4D particle tracking, the ultimate goal of this production.

Due to the charge sharing among pads and thanks to the fact that both peak attenuation and delay are functions of the distance between the hit point and the surrounding pads, the most powerful way to exploit the information given by the RSD is to acquire data from more channels at the same time -- let's say, from a cluster of pads -- and manage them with an analogical scheme. Indeed, we observed that the best performances of our detectors in reconstructing the position and time of tracks can be obtained -- at run-time or offline -- combining all the signals of the cluster by means of amplitude-weighted centroids (for further details on this method, see Refs.~\cite{2019Tornago_35thRD50,2019Arcidiacono_HSTD}).

Before the reconstruction process, each acquisition must be calibrated. Indeed, being the signal amplitude so crucial, it is of primary importance to get rid of any eventual imbalances in the readout chains (wires, amplifiers, etc) that might give rise to misleading results. To this aim, it is enough to set a minimum cluster of 4 pads, hit the laser in the geometrical center of the cluster and then to calibrate each channel so that all the peaks are equal in time and amplitude. But since the broadening introduced by the delay of peaks would degrade the final time resolution, also all the other signals, whatever the hit position, must be \mbox{re-aligned} when estimating the timing performances. To do this, it is necessary to characterize the RSD under study by extracting the peak delay versus position curve (as the one reported in panel (c) of Figure~\ref{fig:signal_delay}) and apply a proper \mbox{distance-dependent} offset to each reconstructed time of arrival.

\begin{figure}[!t]\begin{center}
\includegraphics[width=\columnwidth]{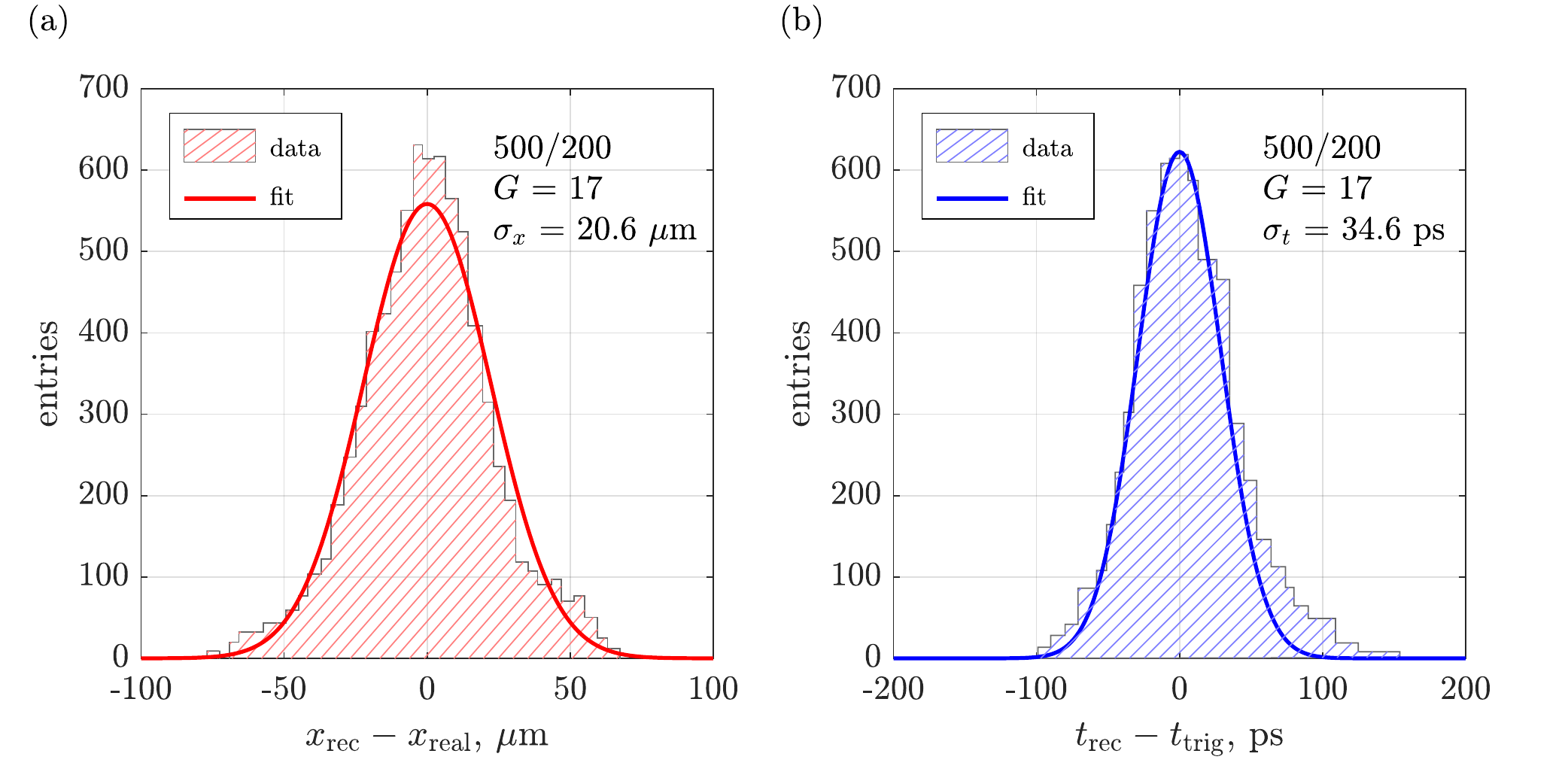}
\caption{Statistical distributions representing the difference of reconstructed position $x_\text{rec}$ (a) and time $t_\text{rec}$ (b) with respect to their reference values (respectively, the real hit point position $x_\text{real}$ and the trigger time $t_\text{trig}$) measured with a pulsed IR laser (1 MIP) of the TCT setup in a \mbox{500~$\mu$m-pitch} RSD1 sample at gain 17. The values $\sigma_x$ and $\sigma_t$, respectively the space and time resolution of the reconstructed data, correspond to the standard deviation of the gaussian curve (solid line) fitting the data (histogram).}\label{fig:resolution1}
\end{center}\end{figure}

\begin{figure}[!b]\begin{center}
\includegraphics[width=\columnwidth]{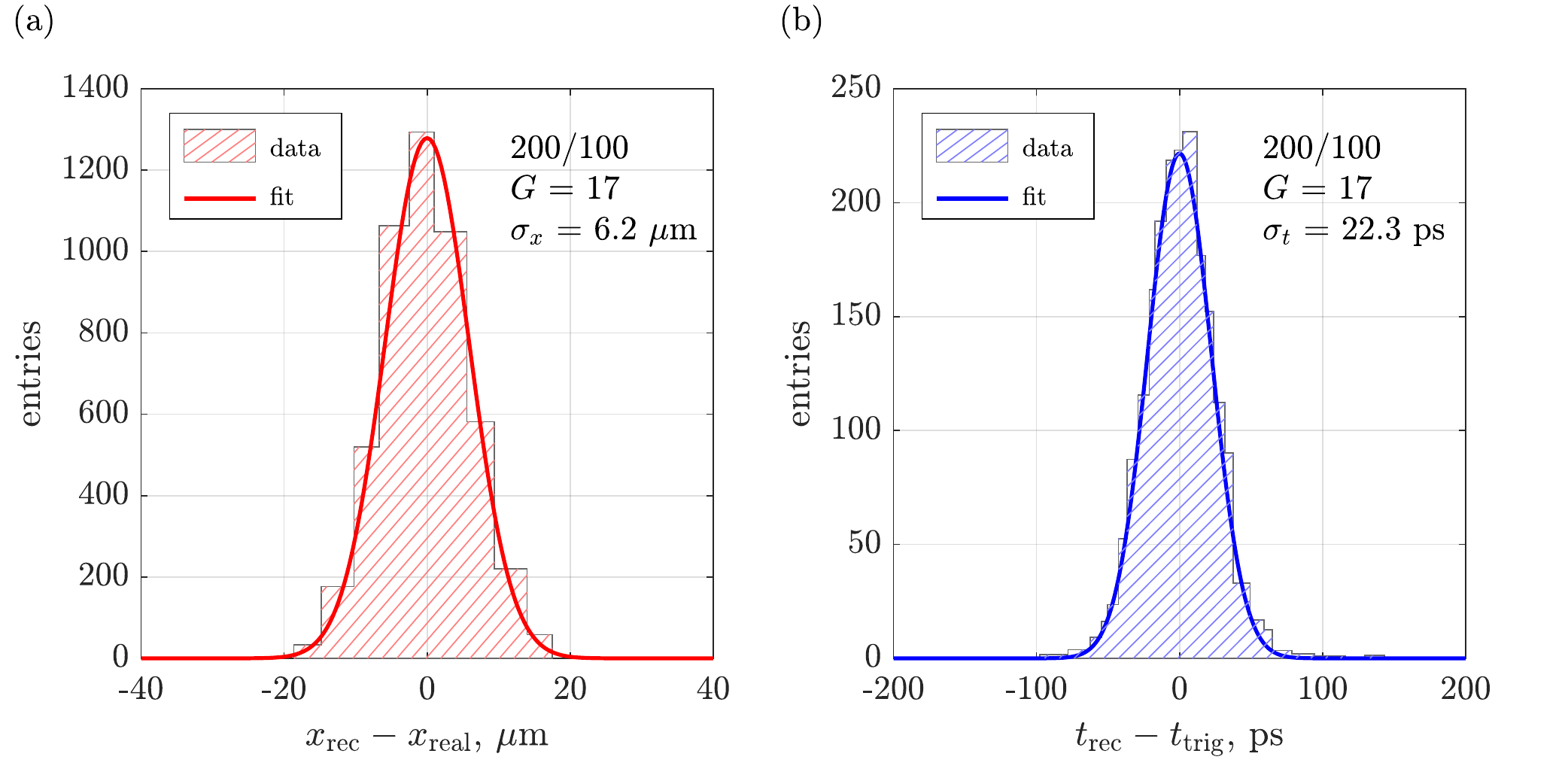}
\caption{Same plots of the previous Figure for the case of a \mbox{200~$\mu$m-pitch} RSD1 sample at gain 17.}\label{fig:resolution2}
\end{center}\end{figure}

Once the \mbox{pre-calibration} has been implemented and the detector is characterized under the signal propagation standpoint, it is possible to use the TCT pulsed laser to generate a given amount of AC signals in several positions of the spot on the detector surface. Acquiring a large number of events and obtaining the corresponding centroids for position and time for each hit point it is also possible to calculate the distributions of their difference with respect to the reference values, that are: (\emph{i}) the exact position of the TCT moving stage $x_\text{real}$, for what concerns the space reconstruction, and (\emph{ii}) the trigger time $t_\text{trig}$, given by a photodiode placed within the optical path between the laser and the detector, regarding the time reconstruction.

Figure~\ref{fig:resolution1} and \ref{fig:resolution2} show the calculated (histogram) and fitted (gaussian) distributions of the reconstructed position and time (respectively $x_\text{rec}$ and $t_\text{rec}$) with respect to their references, calculated by acquiring from a cluster of 4 pads on \mbox{200~$\mu$m-pitch} and \mbox{500~$\mu$m-pitch} RSD1 samples, and selecting only the hit positions giving good signals (i.e., all amplitudes $>$10~mV). The distributions come from the sum of the events acquired in all the hit points where the laser was shot. As one may see, for a gain 17 and 1 MIP laser, both the 500/200 and 200/100 structure have a space resolution $\sigma_x$ better than the nominal distance between pads and a very good time resolution $\sigma_t$, which is scaling with the detector pitch.

Notice that the results are very preliminary because the centroid method has to be improved and due to the use of only small clusters. Moreover, the resolutions are slightly degraded by the operation of convolution of all the reconstructed hit points (not shown here), that in general perform even better. So more analysis has to be done in order to find the ultimate intrinsic performances of the RSD1 production, that we expect to be higher than what presented here.

To further investigate the trend of $\sigma_x$ and $\sigma_t$ as a function of the geometry configuration and of the bias, we repeated the same experiment with other structures, operating the RSD1 under study at different values of gain. Some results of such testing campaign are reported in Table~\ref{tab:resolutions}.

\begin{table}[!h]\begin{center}
\begin{tabular}{|c|c|c|c|c|c|c|}
\hline
\multirow{2}{*}{RSD1} & \multicolumn{2}{c|}{\textbf{Gain 12}} & \multicolumn{2}{c|}{\textbf{Gain 17}} & \multicolumn{2}{c|}{\textbf{Gain 24}} \\ \cline{2-7} 
 & $\sigma_x$ [$\mu$m] & $\sigma_t$ [ps] & $\sigma_x$ [$\mu$m] & $\sigma_t$ [ps] & $\sigma_x$ [$\mu$m] & $\sigma_t$ [ps] \\ \hline\textbf{100/70} & 3.2 & 17.6 & 2.8 & 15.3 & 2.5 & 13.9 \\ \hline
\textbf{200/100} & 8.6 & 31.1 & 6.2 & 22.3 & - & - \\ \hline
\textbf{200/190} & 17.9 & 58.7 & 14.3 & 62.6 & 8.8 & 59.9 \\ \hline
\textbf{500/200} & 27.3 & 45.7 & 20.6 & 34.6 & 20.6 & 32.6 \\ \hline
\end{tabular}
\caption{Summary of the most relevant measurements of concurrent space and time resolution in some RSD1 samples. The results are obtained by shooting the TCT laser on different positions of a cluster of 4 pads, combining the waveforms to find the reconstructed position and time of arrival as described in the text and summing up the events coming from all the hits.}\label{tab:resolutions}
\end{center}\end{table}

The table essentially shows that both the space and time resolution improve with high gain and small pad pitch and side. In fact, the structure 100/70 at gain 24 resulted to be the most performing, with a \mbox{micron-level} space resolution and -- contemporarily -- an outstanding timing \mbox{figure-of-merit} of less than 14~ps, probably the most advanced combination of properties ever realized in a single Silicon tracker, at the best of our knowledge.

\section{Large-area tracking with RSD}

One of the most challenging goals of the RSD technology, besides the radiation hardness, is the realization of continuous and homogeneous multiplication and resistive layers. Apart from the practical difficulties, mostly linked to the implantation processes, the \mbox{high-precision} \mbox{4D-tracking} requires uniformity also in the performance, among which the multiplication, discharge slowdown and, above all, the RC characteristics of the sensor.

To test the robustness of our detectors under the standpoint of operability on a large area, we have chosen a device from the RSD1 production as the ideal workbench. The detector consists of a 64$\times$64 pixel matrix having pitch 50~$\mu$m and pad side 45~$\mu$m. Originally designed to be \mbox{flip-chipped} onto CHIPX65, the Torino version of the RD53A ASIC~\cite{RD53A_chip}, with its \mbox{$\sim$11~mm$^2$} it represents the largest matrix detector of the entire RSD1 batch.

Three clusters of pads have been identified: cluster 1 is composed by the 4 pads at the corners of the second innermost square around the center of the matrix, while both clusters 2 and 3 consist of similar groups of 4 pads but located, respectively, in the middle of one edge and at corner of the matrix. Due to the high difficulty in \mbox{wire-bonding} so small pitched structures, only seven connections survived: the whole cluster 1 plus two pads from the second and only one from the last.

\begin{figure}[!h]\begin{center}
\includegraphics[width=\columnwidth]{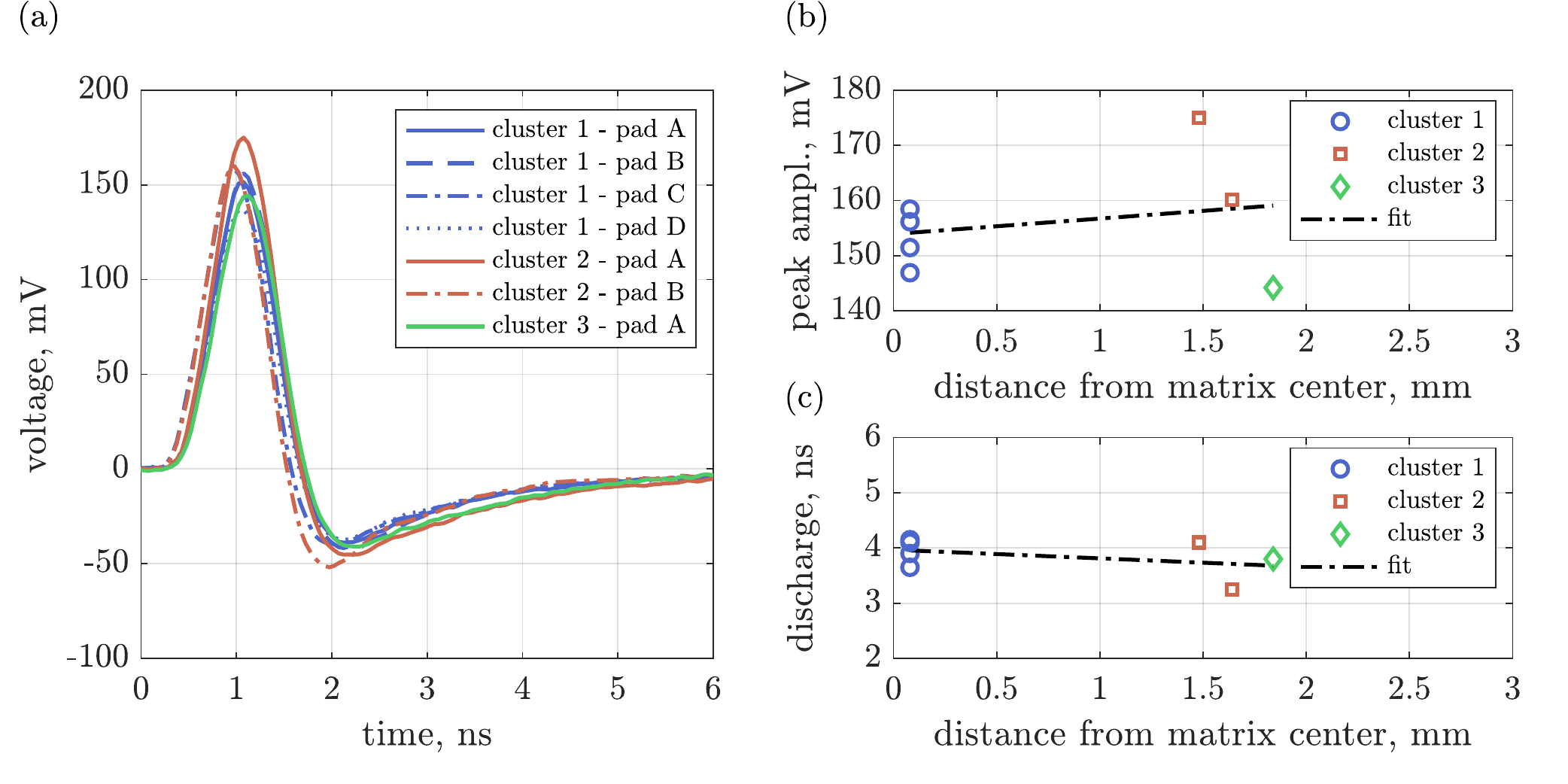}
\caption{Collection of signals (average of several events) acquired at different positions of a \mbox{large-area} RSD1 matrix (a) with the corresponding dispersion of peak amplitude (b) and discharge time (d) as functions of the distance from the considered pad to the center of the detector matrix.}\label{fig:largearea}
\end{center}\end{figure}

The waveforms (averages of several events) reported in panel (a) of Figure~\ref{fig:largearea} are obtained shining the 1 MIP pulsed laser of the TCT setup close to each of the seven pads. Since the signals are not perfectly overlapped, we studied the trend of gain and RC characteristic as a function of the position of the cluster within the matrix and, in particular, of the distance of each pad from the geometrical center of the detector. As an indication of the gain we plotted in panel (b) the maximum amplitude while to check the RC properties we monitored, in panel (c), the discharge time of each waveform.

As we may see, there is more dispersion of data within each cluster -- regarding both peak amplitude and discharge time -- than among different clusters. This can be easily explained by the fact that it is clearly challenging to \mbox{wire-bond} pads with a 50/45~$\mu$m configuration. Indeed, an influence of the wedge footprints may occur, hiding part of the non metallized Silicon, already limited in a sensor with an interpad of just 5~$\mu$m. This locally reduces the penetrating power of the laser, whose spot is approximately 15~$\mu$m wide. Moreover, some non negligible capacitive or inductive effects may rise just from the presence of wires, given the small dimension of pads. In conclusion, as demonstrated by the fit lines, the slight differences among the considered waveforms can be ascribed to small perturbations of the system and there is no correlation between the position within the matrix and the signal formation, even at large distance.


\section{Comments and conclusions}

RSD are innovative detectors for \mbox{4D-tracking} developed at INFN Torino and manufactured at FBK, Trento, which are based on the internal gain mechanism. They are an evolution of the standard LGAD technology because they get rid of any isolation implant, which deteriorates the geometrical acceptance and affect the overall detector efficiency. To obtain concurrent high performances on the spatial and time tracking without segmentation structures RSD introduce a new readout logic based on the resistive \mbox{AC-coupled} scheme, where the multiplied charges are slowed down in a resistive sheet and then, thanks to a dielectric buffer layer, induced via capacitive coupling in the metal pads located on top of the detector.

In this work we presented a full characterization of a first batch, named RSD1, starting with the preliminary electrical tests performed to verify the production quality (see also Ref~\cite{2019Mandurrino_VERTEX}). Thanks to $I(V)$ and $C(V)$ static characterization we determined a great uniformity of the most important technological aspects of RSD1, both within each wafer and among the wafers based on the same process parameters. Then we described the signal generation as well as the dynamic properties of our detectors, thanks to an extensive testing campaign carried out with a TCT infrared laser.

After demonstrating that RSD1 succeeded in the challenging goal of realizing the 100\% \mbox{fill-factor} (i.e., the ratio between the active and the total area of a detector), even in \mbox{very-fine} pitch RSD, we also characterized this first production under the \mbox{4D-tracking} standpoint. Indeed, measurements at the TCT revealed that the intrinsic spatial resolution is highly better than the nominal pitch and that both space and time precision (\emph{i}) scale with the pad pitch and side and (\emph{ii}) improve when operating at high gain values. Carrying out a testing campaign on several structures we found that the most performing RSD1 sample, a \mbox{100~$\mu$m-pitch} square matrix with 70~$\mu$m pad side, has concurrently shown \mbox{$\sigma_x=$2.5~$\mu$m} and \mbox{$\sigma_t=$13.9~ps}, which probably makes RSD1 the first detectors measuring both space and time with such unprecedented precision.

Moreover -- with the characterization of a large RSD1 matrix composed by 50/45~$\mu$m 64$\times$64 pixels -- also the uniformity of signal generation has been demonstrated, showing that neither the multiplication nor the RC properties are correlated with the position of the measured pads.

In conclusion, we have produced and characterized the run RSD1, which shows the great potentialities of such technology in view of the 4D particle tracking. Although some results are still preliminary (we need to improve the reconstruction algorithms) and the characterizations of the irradiated samples are still ongoing, the first outcomes about space and time resolution are extremely good. Thus, we are currently working to find the intrinsic limits in the performances of our detectors, that we expect may be even better than the results shown here.

\section*{Acknowledgements}

This work and the RSD project are part of the collaboration framework between INFN and FBK, and are supported by INFN \mbox{Gruppo-V} through the ``2017 Young Researchers Grant'' funding program (as in the INFN announcement No.19105 and the related deliberation No.19567).
The work was also partially funded by Horizon2020 Grants no. UFSD669529 and 654168 (\mbox{AIDA-2020}), the U.S. Department of Energy Grant no. \mbox{DE-SC0010107}, Dipartimenti di Eccellenza and Universit\`{a} di Torino (ex L. 232/2016, art. 1, cc. 314, 337).
The authors would also thank the RD50 Collaboration at CERN for the scientific support.


\end{document}